# Nonmonotonic Evolution of the Superconducting Transition Temperature and Robust Multigap Extended *s*-wave + *s*-wave Pairing in Zn-Substituted FeSe Single Crystals

Han-Shu Xu[1,2*], Changhao Ding[3,4], Guanyin Gao[2], Xin Zhang[1], Xinyu Yin[1], Xucai Kan[5], Jiaping Hu[2], Wen Xie[2], Wensen Wei[6], Yuxiao Hou[7], Keyu An[8,9*], Haoxiang Li[8], Kaibin Tang[2,10*], and Yu-Yan Han[6*]

[1] Department of Applied Physics, Anhui Medical University, Hefei 230032, China

[2] Hefei National Research Center for Physical Sciences at the Microscale, University of Science and Technology of China, Hefei 230026, China

[3] Materials Sciences Division, Lawrence Berkeley National Lab, Berkeley, CA, 94720, USA

[4] Department of Materials Science and Engineering, University of California, Berkeley, CA, 94720, USA

[5] Engineering Technology Research Center of Magnetic Materials, School of Physics & Materials Science, Anhui University, Hefei 230601, China

[6] Anhui Key Laboratory of Low-Energy Quantum Materials and Devices, High Magnetic Field Laboratory, Chinese Academy of Sciences, Hefei 230031, China

[7] Materials Characterization and Preparation Facility, The Hong Kong University of Science and Technology (Guangzhou), Guangzhou 511453, China

[8] Advanced Materials Thrust, Function Hub, The Hong Kong University of Science and Technology (Guangzhou), Guangzhou 511453, China

[9] Institute of Applied Physics and Materials Engineering, University of Macau, Macao SAR 999078, China.

[10] Department of Chemistry, University of Science and Technology of China, Hefei 230026, China

## Abstract

We report a systematic study of superconductivity on $Fe_{1-x}Zn_xSe$ single crystals synthesized over a broad Zn doping range ($x$ = 0-0.023). High-quality single crystals across all compositions range exhibit superconducting transitions, while the transition temperature $T_c$ shows a pronounced nonmonotonic dependence on Zn doping concentration, indicating that the underlying mechanism govering $T_c$ its evolution cannot be explained solely by simple impurity pair breaking alone. Magnetization and transport measurements confirm the bulk behavior of superconductivity and reveal enhanced scattering effects with Zn doping. Low-temperature specific heat is consistently described by a two-gap scenario composed of an isotropic *s*-wave gap and an anisotropic extended *s*-wave gap, whereas single-gap and alternative pairing symmetries fail to describe the data. The nearly unchanged relative weights of the two gap components suggest the weak interband scattering induced by Zn substitution, thereby preserving multiband superconductivity. These results demonstrate the robustness of multigap superconductivity in FeSe and impose stringent constraints on candidate pairing mechanisms, highlighting the role of multiband electronic structure and anisotropic gap formation.

## 1. Introduction

Since its discovery, FeSe has attracted extensive attention due to its unique properties. On the one hand, FeSe possesses a relatively simple crystal structure (space group $P4/nmm$), making it an ideal platform for investigating the intrinsic physics of iron-based superconductors [1,2]. On the other hand, its $T_c$ is highly sensitive to external tuning. For instance, intercalation of organic molecules, co-intercalation of alkali or alkaline-earth metals with organic species, and the application of external hydrostatic pressure can significantly enhance $T_c$, reaching values as high as 30-45 K [3-6]. Notably, under high pressure, $T_c$ exhibits a nonmonotonic evolution and multiple superconducting phases (e.g., double-dome behavior) emergence, often accompanied by competing magnetic or structural orders [7-9]. Despite extensive studies, the superconducting pairing mechanism in FeSe remains under debate. Most specific heat measurements on bulk single crystals suggest a nodeless superconducting gap with pronounced anisotropy [10-13]. However, ultralow-temperature thermal conductivity measurements have yielded conflicting conclusions regarding the presence of nodes, and no consensus has not yet been achieved [14-16]. In addition, spectroscopic studies, such as scanning tunneling microscopy and angle-resolved photoemission spectroscopy, tend to support the existence of nodes or deep gap minima [17-19]. Overall, the pairing symmetry and gap structure in FeSe remain unresolved, calling for further systematic investigations.

Nonmagnetic impurity substitution is a key approach for probing superconducting pairing mechanisms. In conventional isotropic $s$-wave superconductors, the Anderson theorem indicates that nonmagnetic impurities have little effect on Cooper pairing and thus exert minimal influence on $T_c$ [20]. In contrast, for sign-reversing states such as $d$-wave or $s\pm$-wave, nonmagnetic impurities induce interband scattering, which acts as an effective pair breaker and consequently suppresses superconductivity [21-23]. Magnetic impurities, however, strongly suppress superconductivity in most pairing symmetries by breaking Cooper pairs, as described by the Abrikosov-Gorkov theory [24]. In FeSe system, substitution studies have mainly focused on the Se site, such as Te or S substitution. The FeSe$_{1-x}$Te$_x$ system (near $x \approx 0.5$) has been reported to exhibit possible topological superconductivity with signatures of Majorana zero modes [25]. In contrast, Fe-site substitution, especially by nonmagnetic elements, remains less explored. Cu doping rapidly suppresses superconductivity in Fe$_{1-x}$Cu$_x$Se and induces an Anderson localization-driven metal-insulator transition [26-28]. For Fe$_{1-x}$Co$_x$Se, $T_c$ decreases with increasing Co content and vanishes at $x \approx 0.036$. However, whether Co$^{2+}$ should be regarded as a purely nonmagnetic impurity remains debated [29,30], as high-spin Co$^{2+}$ may carry a local magnetic moment.

In this work, we employed nonmagnetic Zn substitution at the Fe site in FeSe and successfully synthesized a series of high-quality Fe$_{1-x}$Zn$_x$Se single crystals, where Zn substitution has not been previously explored in single crystal system. The superconducting properties were systematically investigated using magnetization, electrical transport, and low-temperature specific heat measurements. The results show that $T_c$ exhibits a non-monotonic evolution with increasing Zn content (Fig. 1). The low-temperature specific heat can be consistently analyzed within a two-gap scenario composed of an anisotropic extended $s$-wave gap and an isotropic $s$-wave gap, in agreement with the general multiband superconductivity picture of FeSe-based systems. Moreover, the nearly invariant relative contributions of the two gap components across different Zn concentrations suggest that Zn substitution only introduces weak interband scattering, thereby preserving the stability of the multigap superconducting state. This behavior is indicative of a superconducting order parameter that is likely sign-preserving. Overall, these findings highlight the robustness of multigap superconductivity in FeSe and place important constraints on the possible pairing mechanisms, emphasizing the essential role of multiband electronic structure and anisotropic gap formation.

## 2. Experimental section and Characterization

In the crystal growth process of FeSe single crystals, 0.62 g of reduced iron powder (99.99%, Alfa Aesar) and 0.78 g of selenium grains (99.999%, Alfa Aesar) were first weighed and loaded into a quartz tube with an outer diameter of 15 mm and a wall thickness of 1.5 mm. Subsequently, 6.22 g $AlCl_3$ (99.9%, Aladdin) and 1.73 g KCl (99.99%, Aladdin) were added as fluxes to facilitate mass transport and the kinetics of crystal growth. After loading, the quartz tube was evacuated to approximately $10^{-3}$ Pa and then sealed under vacuum. The crystal growth was carried out in a two-zone tube furnace, where the high-temperature zone was maintained at 450 °C and the low-temperature zone at 350 °C, establishing a stable temperature gradient to drive the transport and crystallization process. After approximately 45 days of slow growth, millimeter-sized plate-like FeSe single crystals were successfully obtained. The growth procedure for Zn-substituted $Fe_{1-x}Zn_xSe$ single crystals is identical to that used for the synthesis of FeSe.

X-ray diffraction (XRD) measurements on $Fe_{1-x}Zn_xSe$ single crystals were conducted at room temperature using an X′Pert3 MRD diffractometer with Cu Kα radiation ($\lambda$ = 1.541841 Å) over a 2θ range of 10°-80°. The elemental composition was determined by energy-dispersive spectroscopy integrated into a scanning electron microscope (SEM-EDS, Oxford Ultim Max Infinity 100). The sample thickness used for the estimation of the critical current density was measured using an Olympus BX51M optical microscope. Magnetic properties were measured with the vibrating sample magnetometer option of a Quantum Design Physical Property Measurement System (PPMS). Electrical transport measurements were performed on $Fe_{1-x}Zn_xSe$ single crystals using a standard four-probe configuration, while the specific heat was obtained via the thermal relaxation technique. Both measurements were carried out on the same PPMS. It should be noted that all magnetic-field-dependent measurements reported in this study were conducted with the magnetic fields applied along the *c*-axis (perpendicular to *ab*-plane).

## 3. Experimental Results and Discussions

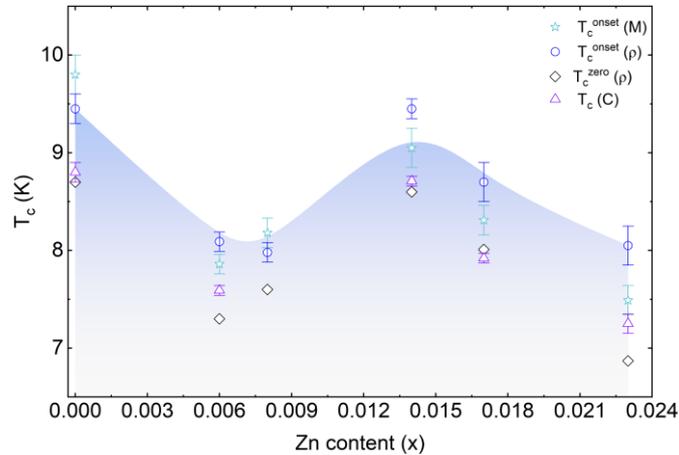

Fig. 1 Phase diagram of Zn-substituted FeSe single crystals, $Fe_{1-x}Zn_xSe$, as a function of Zn

concentration $x$. The $T_c$ are determined from magnetization (M), resistivity ($\rho$), and specific heat (C) measurements.

To verify the crystallinity of Fe$_{1-x}$Zn$_x$Se single crystals with different Zn contents, XRD measurements were performed on their naturally cleaved *ab*-plane. As shown in Fig. 2(a-f), only sharp (00*l*) reflections are observed for all samples, without detectable impurity peaks, indicating the high crystallinity of the synthesized single crystals. Because the FeSe layers are coupled by weak van der Waals interactions, the interlayer spacing $d$, equal to the lattice parameter *c*-axis, is highly sensitive to elemental substitution, organic molecular intercalation, and external pressure. Based on the XRD results, the lattice parameter $c$ exhibits a non-monotonic evolution with increasing Zn content, remaining overall close to about 5.50 Å, consistent with the behavior reported in Cr-substitution Cs(V$_{1-x}$Cr$_x$)$_3$Sb$_5$ [31]. This behavior indicates that Zn predominantly substitutes at the Fe sites rather than occupying interstitial positions within the FeSe layers. Fig. 2(g-l) display photographs of Fe$_{1-x}$Zn$_x$Se single crystals with different Zn contents, and the typical dimensions of the single crystals are approximately (2-4)×(1-2)×(0.15-0.3) mm$^3$. To further quantitatively determine the Zn content in the Fe$_{1-x}$Zn$_x$Se single crystals, EDS mapping was performed on millimeter-sized samples. Taking ($x = 0.017$) as a representative example (Fig. 2(m-r) and Fig. 2(t)), the elemental maps reveal a homogeneous distribution of Fe, Zn, and Se throughout the crystal, with clearly identifiable characteristic peaks of Zn.

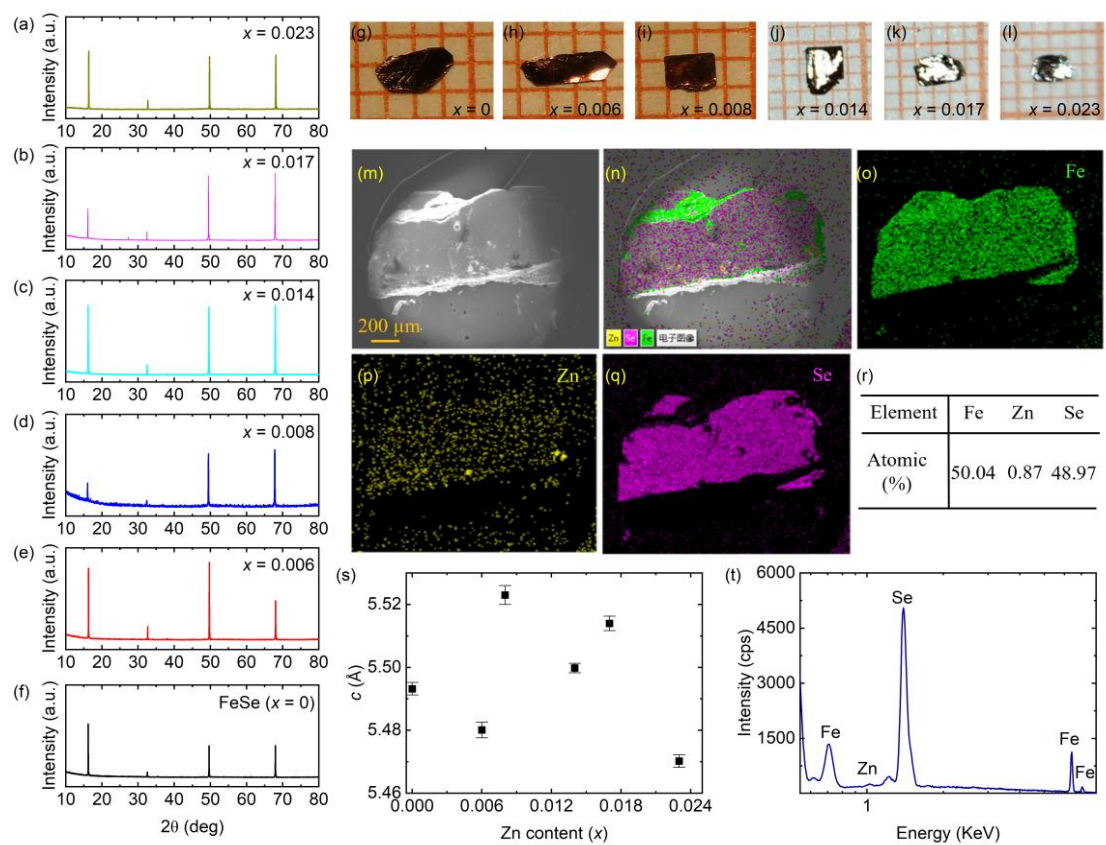

Fig. 2 Elemental composition and structural characterizations of Fe$_{1-x}$Zn$_x$Se single crystals. (a-f) XRD patterns measured on the *ab*-plane of Fe$_{1-x}$Zn$_x$Se single crystals with different Zn contents. (g-l) Photographs of the corresponding single crystals, with each small grid representing 1 mm. (m-q)

Elemental mapping of representative $Fe_{0.983}Zn_{0.017}Se$ single crystals, and (r) the corresponding elemental compositions. (s) Evolution of the *c*-axis lattice parameter as a function of Zn content. (t) EDS spectrum acquired from a freshly cleaved surface of the $Fe_{0.983}Zn_{0.017}Se$ single crystal.

To investigate the effect of Zn substitution on the superconductivity of FeSe, magnetic measurements were carried out. For the parent FeSe (Fig. 3(a) and Fig. S1), the bifurcation between the ZFC and FC curves occurs at 9.8 K approximately, corresponding to an onset superconducting transition temperature $T_c^{onset}$, slightly higher than previous reports [1,10,16]. Furthermore, the FC susceptibility remains close to zero, consistent with strong flux pinning in the superconducting state. Combined with the absence of anomalous magnetic signals, this indicates that the samples are free of detectable ferromagnetic or ferrimagnetic impurities and possess high chemical purity. The magnetization at 1.8 K is about -1.13 emu/g, which corresponds to an apparent superconducting shielding fraction exceeding 100% due to demagnetization effects associated with the sample geometry [13], indicating bulk superconductivity in the FeSe single crystal.

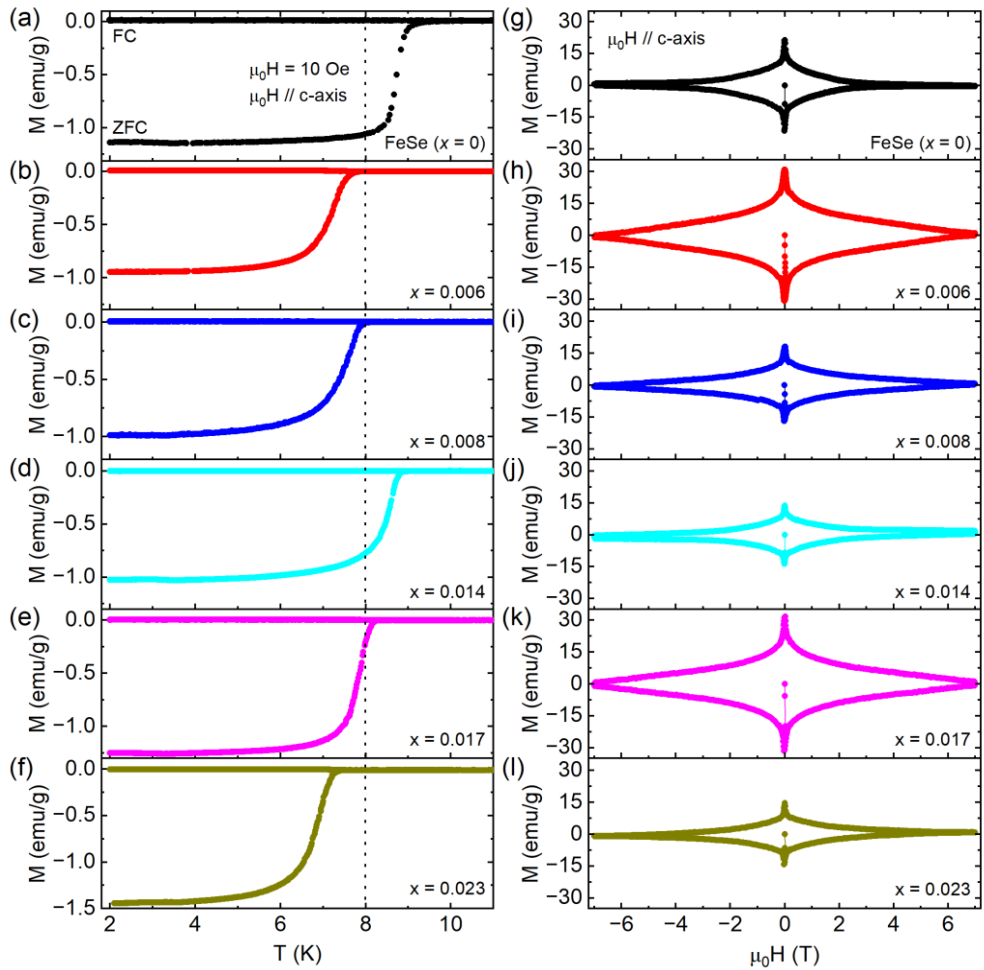

Fig. 3 Magnetic properties of $Fe_{1-x}Zn_xSe$ single crystals measured under an applied field of 10 Oe along the *c*-axis. (a-f) ZFC and FC magnetization. (g-l) Magnetization hysteresis loops measured at 2 K with the magnetic field swept from 0 T to 7 T, then to -7 T, and back to 7 T.

Importantly, in the $Fe_{1-x}Zn_xSe$ single-crystal system, the $T_c^{onset}$ exhibits a non-monotonic dependence on Zn content, as shown in Fig. 3(b-f) and Fig. S1. Upon slight Zn substitution ($x$ = 0.006), $T_c^{onset}$ decreases to 7.86 K. With further increase to $x$ = 0.008, $T_c^{onset}$ recovers to 8.18 K and reaches 9.05 K at $x$ = 0.014. However, with additional Zn incorporation, $T_c^{onset}$ is suppressed again to 8.31 K for $x$ = 0.017 and 7.49 K for $x$ = 0.023. Notably, with increasing nonmagnetic Zn substitution, superconductivity is initially suppressed, then partially recovers, and subsequently suppressed again, a behavior reminiscent of nonmagnetic Ti substitution in $CsV_3Sb_5$, in contrast to the monotonic suppression observed in FeSe with magnetic Co substitution [32,33]. In addition, the ZFC magnetization indicates that $Fe_{1-x}Zn_xSe$ retains bulk superconductivity throughout the entire substitution range of $x$ = 0-0.023. We further increased the Zn content ($x$ = 0.03), the single crystal still exhibits good crystallinity (Fig. S2(a)), but the superconducting shielding fraction decreases rapidly, indicating a deviation from bulk superconductivity (Fig. S2(b)). Therefore, higher Zn concentrations are not considered in the present study.

**Table I**. Superconducting parameters of $Fe_{1-x}Zn_xSe$ single crystals, including the $H_{c1}$, $H_{c2}$, $J_c$ at 2 K in the self-field, $\lambda$, $\xi$, and $\kappa$.

| $Fe_{1-x}Zn_xSe$ | $H_{c1}$ (Oe) | $H_{c2}$ (T) | $J_c$ ($\times 10^4$ A/cm$^2$) | $\lambda$ (nm) | $\xi$ (nm) | $\kappa$ |
|---|---|---|---|---|---|---|
| $x$ = 0 | 65 | 16.28 | 8.09 | 329 | 4.50 | 73.35 |
| $x$ = 0.006 | 45 | 13.01 | 8.53 | 400 | 5.03 | 79.54 |
| $x$ = 0.008 | 30 | 15.03 | 10.04 | 507 | 4.68 | 108.32 |
| $x$ = 0.014 | 39 | 15.49 | 6.06 | 438 | 4.61 | 95.09 |
| $x$ = 0.017 | 50 | 13.99 | 19.30 | 402 | 4.85 | 82.89 |
| $x$ = 0.023 | 19 | 12.56 | 6.78 | 647 | 5.12 | 126.52 |

As shown in Fig. 3(g-l), the $M$(H) curves of $Fe_{1-x}Zn_xSe$ single crystals with different Zn contents exhibit the characteristic behavior of typical type-II superconductors. With increasing Zn content, the width of the $M$(H) hysteresis loop, $\Delta M$, defined as the difference between the magnetization in the increasing- and decreasing-field branches at the same applied field, $\Delta M = M_{up} - M_{down}$, varies, reflecting the modulation of flux pinning strength and critical current density $J_c$. Furthermore, the $J_c$ ($\mu_0 H$ // $c$-axis) is estimated using the Bean critical state model, $J_c = 20\Delta M/[a(1 - a/3b)] \approx 20\Delta M/a$, where $a$ is the sample thickness along the $c$-axis and $b$ represents the in-plane dimension of the sample, with the approximation valid for $a << b$ [34]. The $J_c$ at 2 K in the self-field is summarized in Table I. The lower critical field $H_{c1}$ was estimated by fitting the linear magnetic response in the low-field region (Meissner state) and determining the field at which the magnetization deviates from this linear behavior (Fig. S3). In further, based on the Ginzburg-Landau relations $\xi = \sqrt{\frac{\Phi_0}{2\pi H_{c2}}}$, $H_{c1} = \frac{\Phi_0}{4\pi \lambda^2} \ln(\frac{\lambda}{\xi})$, and $\kappa = \frac{\lambda}{\xi}$, London penetration depth $\lambda$, coherence length $\xi$, and Ginzburg-Landau parameter $\kappa$ are determined (Table I) [34]. Here, $\Phi_0$ is the magnetic flux quantum with a value of 2.07 $\times$ 10$^{-15}$ Wb, and $H_{c2}$ is the upper critical field obtained from fitting using the Werthamer-Helfand-Hohenberg (WHH) theory for orbital-limited superconductors [35], as discussed later. In short, Table I shows that the superconducting parameters vary nonmonotonically with Zn content, indicating that Zn substitution systematically modifies the superconducting properties.

To further characterize the electrical transport properties of $Fe_{1-x}Zn_xSe$ single crystals, in-plane resistivity measurements within the *ab*-plane were performed (Fig. 4). For the parent FeSe, a kink in the resistivity is observed at approximately $T^* \approx 86.3$ K upon cooling, corresponding to the tetragonal-to-orthorhombic structural transition with nematic order [1,13]. Upon further cooling, superconductivity emerges with an onset temperature of about $T_c^{onset} \approx 9.45$ K. In addition, the residual resistivity ratio (RRR) is about 70, where RRR = $\rho$(300 K = 2.01 mΩ·cm$^{-1}$)/$\rho$(0 K = 0.029 mΩ·cm$^{-1}$), with $\rho$(0 K) estimated by a linearly extrapolation of the normal-state resistivity to zero temperature (Fig. S5), further confirming the high quality of our synthesized FeSe single crystals. With increasing Zn content, the evolution of the $T_c$ in Fe$_{1-x}$Zn$_x$Se is consistent with that obtained from magnetic measurements. Notably, the characteristic temperature $T^*$ follows the same trend, increasing or decreasing in tandem with $T_c$, indicating a close correlation between the two, as shown in Fig. 1. In addition, for the single crystals with $x$ = 0.006, 0.008, 0.014, 0.017, and 0.023, the RRR are 21, 16, 37, 30, and 14, respectively (Fig. S5). These results indicate that, although the Zn-substituted crystals still maintain high crystal quality, the reduction of RRR compared to the parent FeSe can be attributed to the inevitable impurity scattering introduced by Zn substitution.

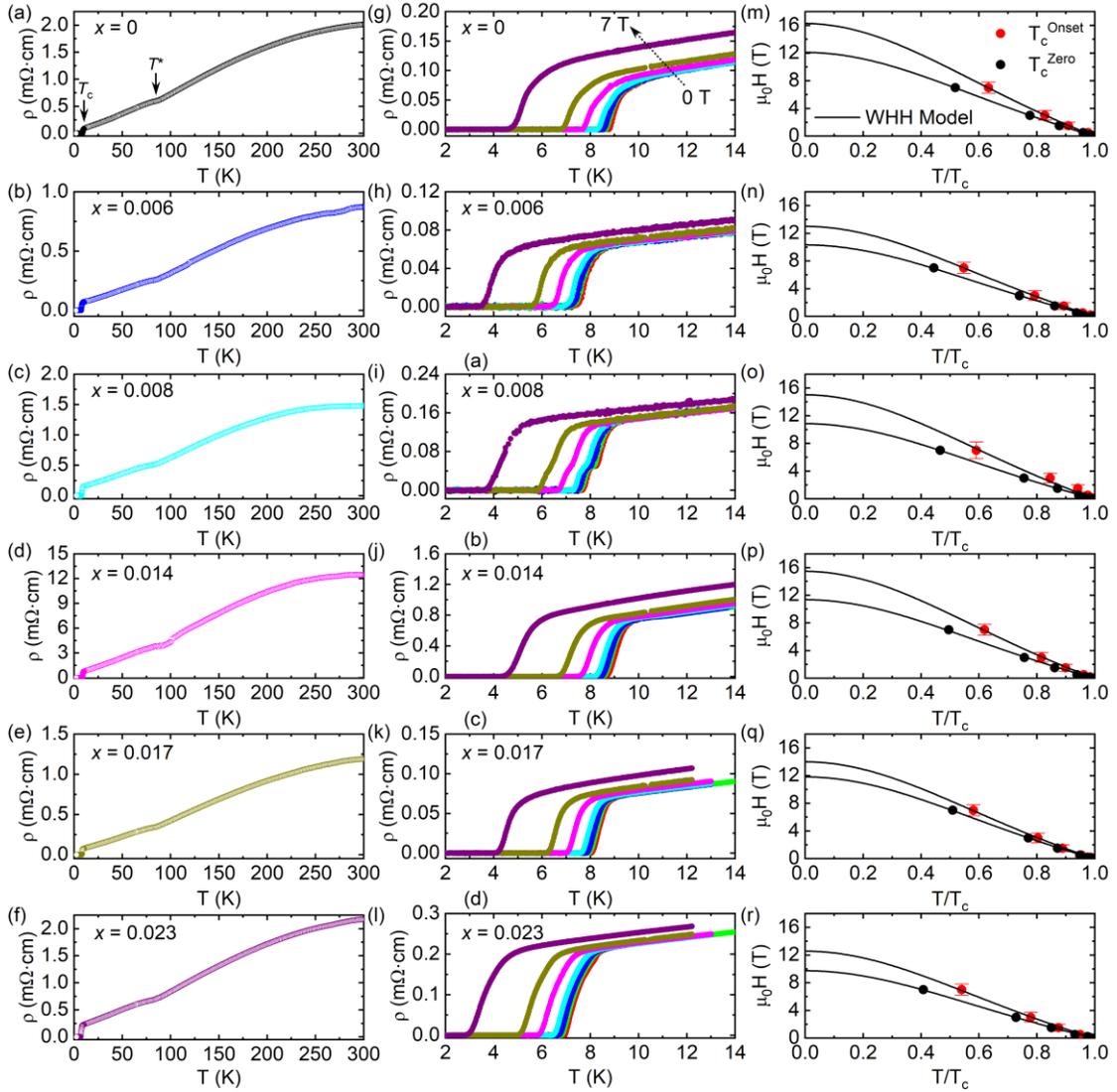

Fig. 4 Electrical transport properties of Fe$_{1-x}$Zn$_x$Se single crystals, measured with current of 3 mA applied in the *ab*-plane and magnetic field parallel to the *c*-axis. (a-f) Temperature-dependent

resistivity of $Fe_{1-x}Zn_xSe$ with different Zn contents measured at zero field in the range of 2-300 K. (g-l) Resistivity in the temperature range of 2-14 K under magnetic fields of 0 T, 0.01 T, 0.1 T, 0.2 T, 0.5 T, 1.5 T, 3.0 T, and 7.0 T. (m-r) Upper critical field as a function of $T/T_c$ for $Fe_{1-x}Zn_xSe$ derived from resistivity measurements. The solid lines represent fits using the WHH model. Here, $T_c^{onset}$ and $T_c^{zero}$ are determined from resistivity, with the criteria defined in Fig. S4.

To investigate the effect of an external magnetic field on superconductivity, the temperature dependence of resistivity in the range of 2-14 K was systematically measured under various applied fields. As shown in Fig. 4(g-l), for $Fe_{1-x}Zn_xSe$ single crystals, the $T_c$ is progressively suppressed with increasing magnetic field, indicating a pronounced suppression of the superconducting state by the applied field. Notably, the upper critical field $H_{c2}$, determined from both $T_c^{onset}$ and $T_c^{zero}$, can be well described by the WHH model (Fig. 4(m-r)). In the fitting, the Maki parameter $\alpha$ and the spin-orbit scattering constant $\lambda_{so}$ are set to zero, indicating that the superconductivity is predominantly governed by the orbital limiting effect. This approximation is commonly adopted when spin-paramagnetic and spin-orbit effects are negligible, as described in the WHH theory [35,36]. The WHH model is generally expressed as:

$$ln\frac{1}{t} = \left(\frac{1}{2} + \frac{i\lambda_{so}}{4\gamma}\right)\psi\left(\frac{1}{2} + \frac{\bar{h} + \frac{\lambda_{so}}{2} + i\gamma}{2t}\right) + \left(\frac{1}{2} - \frac{i\lambda_{so}}{4\gamma}\right) \times \psi\left(\frac{1}{2} + \frac{\bar{h} + \frac{\lambda_{so}}{2} - i\gamma}{2t}\right) - \psi\left(\frac{1}{2}\right)$$

where $t = T/T_c$, $\gamma \equiv \left[(\alpha\bar{h})^2 - (\lambda_{so}/2)^2\right]^{1/2}$, and $\psi$ is the digamma function. The reduced magnetic field $h^*$ is defined as $h^* \equiv \frac{\bar{h}}{\left(-\frac{d\bar{h}}{dt}\right)_{t=1}} = \frac{\pi^2 \bar{h}}{4} = \frac{H_{c2}}{\left(-\frac{dH_{c2}}{dt}\right)_{t=1}}$, which provides a dimensionless scaling of the upper critical field near $T_c$. In the orbital-limited case with $\alpha = 0$ and $\lambda_{so} = 0$, the WHH model reduces to the orbital-limited case, yielding the well-known relation $\mu_0 H_{c2}^*(0) = -0.693 T_c \left(\frac{d\mu_0 H_{c2}}{dT}\right)_{T_c}$. Based on this relation, the zero-temperature upper critical fields for $Fe_{1-x}Zn_xSe$ are estimated and summarized in Table I, where the values are inferred from $T_c^{onset}$.

It is noteworthy that for $x = 0.008$, the zero-field resistivity exhibits a step-like evolution upon cooling, as shown in Fig. 4(i) and Fig. S6. It initially displays a rapid drop at approximately 8.83 K, followed by a plateau-like feature in the range of 8.02-8.25 K. Upon further cooling below 8.02 K, the resistivity drops sharply again and eventually reaches zero at about 7.6 K. Magnetization measurements for the same single crystal reveal that the ZFC and FC curves bifurcate at approximately 8.15 K (Fig. 3(c) and Fig. S1), marking the onset of superconductivity. This $T_c^{onset}$ in magnetization measurement coincides with the onset of the resistivity drop occurs at 8.02 K in transport measurements. In contrast, the initial resistivity decrease at about 8.83 K does not signify a true superconducting transition, but is more plausibly attributed to precursor effects, such as superconducting fluctuations or local pairing, indicating that macroscopic phase coherence has not yet been established at this temperature.

Magnetotransport measurements were performed on $Fe_{1-x}Zn_xSe$ single crystals with $x = 0$, 0.006, and 0.014, as shown in Fig. 5. At 2 K, all three pieces of single crystals remain in the superconducting state and exhibit zero resistance as the magnetic field is increased from 0 T to 7 T (Fig. 5(a)). This behavior indicates that the applied field is insufficient to fully suppress superconductivity at this temperature. At 10 K, where the system is in the normal state, the FeSe exhibits a pronounced positive magnetoresistance, defined as MR = [ρ(H)-ρ(0)/ρ(0)]×100%, reaching approximately 80% at 7 T (Fig. 5(b)). The magnetoresistance shows a clear deviation from a quadratic field dependence in the low-field regime, while an approximately linear dependence of MR on $B^2$ is recovered at higher fields (above 5 T), following MR ≈ 0.0105 $B^2$, where $B = \mu_0 H$ is the magnetic induction. The low-field deviation from the quadratic behavior is attributed to multiband effects and the presence of multiple scattering channels, which are intrinsic to FeSe. Such behavior is commonly observed in iron-based superconductors and reflects the complex electronic structure and carrier dynamics in this system [37-39].

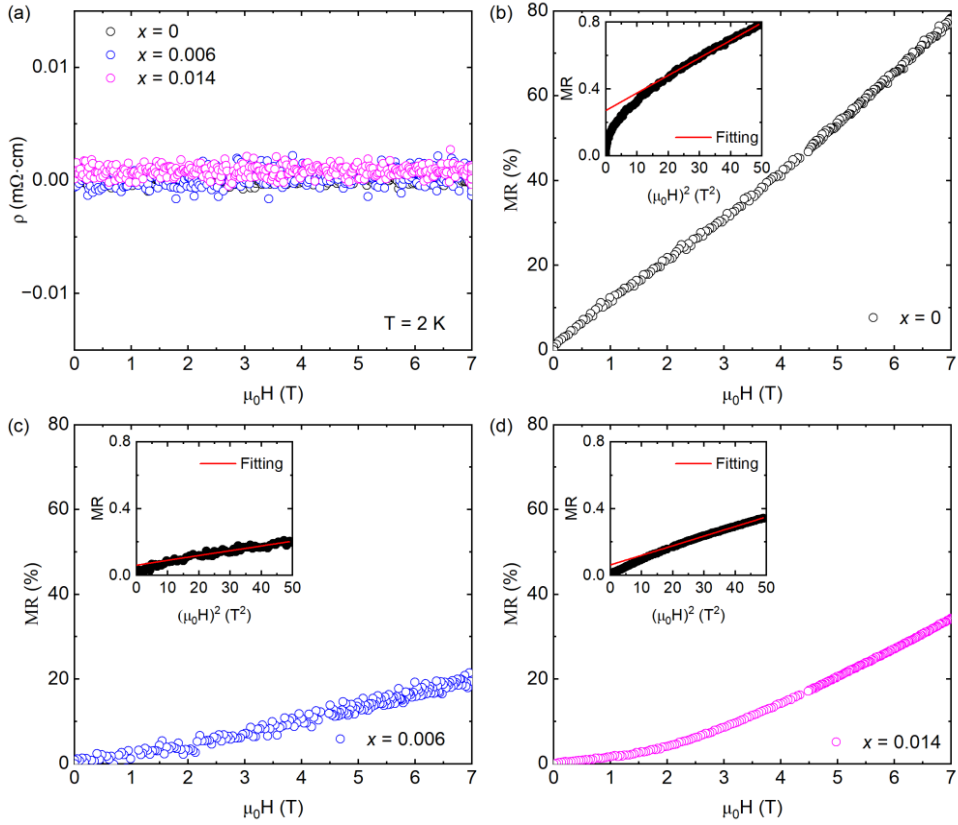

Fig. 5 Magnetotransport properties of $Fe_{1-x}Zn_xSe$ ($x$ = 0, 0.006, and 0.014) single crystals with the magnetic field applied along the $c$-axis. (a) Magnetoresistance obtained at 2 K of the three single crystals in the superconducting state. (b-d) Magnetoresistance obtained at 10 K for $x$ = 0, 0.006, and 0.014, respectively. The insets show MR as a function of magnetic field; the red solid lines represent linear fits to MR = $k$ ($\mu_0 H$)$^2$ + A, where A is a constant.

The quadratic coefficient $k$ = 0.0105 allows an estimate of the effective transport mobility $\mu_{\text{eff}}$ using the relation MR ≈ ($\mu_{\text{eff}} B$)$^2$. This yields an effective mobility $\mu_{\text{eff}}$ ≈ 0.10 T$^{-1}$ (≈ 10$^3$ cm$^2$·V$^{-1}$·s$^{-1}$). This corresponds to a scattering time $\tau = \mu_{\text{eff}} m^*/e$ on the order of (1.1-4.0)×10$^{-12}$ s, where $m^*$ ≈ (2-7)$m_e$ [40]. This timescale indicates reasonably long-lived quasiparticles and coherent charge transport in the normal state, although it should be noted that $\mu_{\text{eff}}$ reflects an average transport scattering rate over multiple bands and scattering channels, rather than a single well-defined disorder-limited scattering time [41,42]. As Zn substitutes for Fe, the $k$ is reduced by nearly one order of magnitude, reaching $k$ = 0.00287 for $x$ = 0.006. Correspondingly, the effective transport mobility and scattering time are estimated to be $\mu_{eff} \approx 5.36\times10^2$ cm$^2\cdot$V$^{-1}\cdot$s$^{-1}$ and $\tau = (0.6\text{-}2.1)\times10^{-12}$ s, respectively. With a further increase of Zn content to $x = 0.014$, the quadratic coefficient increases to $k = 0.00576$, yielding $\mu_{eff} \approx 7.6\times10^2$ cm$^2\cdot$V$^{-1}\cdot$s$^{-1}$, and $\tau = (0.9\text{-}3.0)\times10^{-12}$ s. The reduction of $\mu_{eff}$ and $\tau$ with Zn substitution indicates enhanced impurity scattering and a suppression of carrier mobility, reflecting a crossover from a relatively clean to a more disordered transport regime. Moreover, the nonmonotonic evolution of the $k$ value suggests that Zn substitution not only introduces disorder but also modifies the multiband electronic structure. In particular, the deviation from a simple monotonic behavior implies a redistribution of transport contributions among different Fermi surface pockets, consistent with the compensated semimetallic nature of FeSe and its sensitivity to impurity-induced scattering.

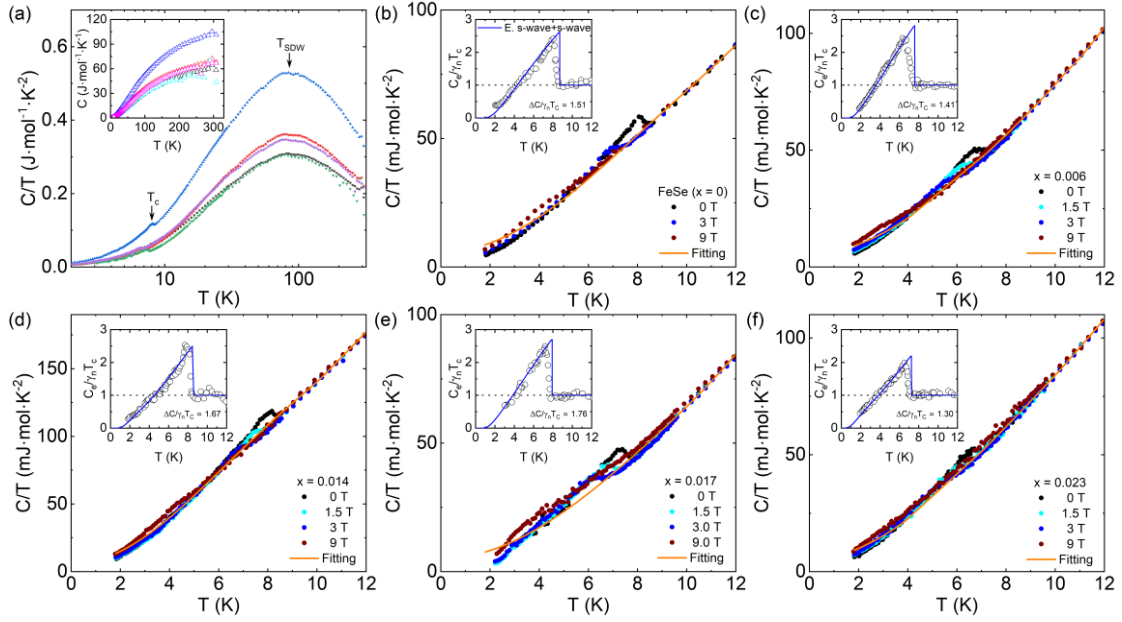

Fig. 6 Specific heat of Fe$_{1-x}$Zn$_x$Se single crystals with the magnetic field applied along the $c$-axis. (a) Temperature dependence of $C/T$ as a function of T in zero field over the range 2-300 K. The inset shows $C$ versus $T$. (b-f) Low-temperature specific heat (1.8-12 K) measured under various magnetic fields for $x = 0$, 0.006, 0.014, 0.017, and 0.023, respectively. The orange solid lines represent the phonon contribution, obtained by fitting the normal-state data under different magnetic fields. The insets display the normalized electronic specific heat $C_e/\gamma_n T$ as a function of temperature, analyzed using a two-gap model with coexisting extended $s$-wave (E. $s$-wave) and conventional $s$-wave components.

The temperature dependence of the specific heat divided by temperature, $C/T$, for Fe$_{1-x}$Zn$_x$Se single crystals is shown in Fig. 6(a). Upon cooling, the single crystals undergo a tetragonal-to-orthorhombic structural transition in the range of 83-88 K, accompanied by the emergence of nematic order, consistent with the resistivity measurements. A clear jump in $C/T$ at 7-9 K is associated with the superconducting transition. Fig. 6(b) shows the temperature dependence of the normalized specific heat of FeSe single crystals in the low-temperature regime under various magnetic fields. With increasing magnetic field, the magnitude of the specific heat jump at the superconducting transition is progressively suppressed and the transition becomes broadened. For example, at 9 T, the sharp anomaly is strongly reduced and evolves into a weak hump around 4 K.

To deconvolute the electronic and lattice contributions to the specific heat and extract the normal-state specific heat parameters, the Debye approximation was employed [43]. Within this framework, the normal-state specific heat $C_n$ can be expressed as the sum of electronic and lattice contributions (orange solid line): $C_n = \gamma_n T + \beta_1 T^3 + \beta_3 T^5 + \beta_5 T^7$, where $\gamma_n T$ represents the electronic specific heat in the normal state, while $\beta_1 T^3 + \beta_3 T^5 + \beta_5 T^7$ accounts for the phonon contribution (Fig. 6(a)). The normal-state specific heat was fitted to obtain the parameters $\gamma_n$ = 5.71 mJ·mol$^{-1}$·K$^{-2}$, $\beta_1$ = 0.75 mJ·mol$^{-1}$·K$^{-2}$, $\beta_3$ = -1.41×10$^{-3}$ mJ·mol$^{-1}$·K$^{-4}$, and $\beta_5$ = 8.34×10$^{-7}$ mJ·mol$^{-1}$·K$^{-6}$, and in further, the Debye temperature $\Theta_D$ can be estimated from the formula $\Theta_D = \left(\frac{12\pi^4 Rn}{5\beta_1}\right)^{1/3}$, where $R$ = 8.314 J·mol$^{-1}$·K$^{-1}$ is the gas constant and $n$ = 2 for FeSe [43], yielding $\Theta_D \approx$ 163 K. The normalized specific heat jump at $T_c$, (i.e., $\Delta C/\gamma_n T_c$ evaluated via entropy conserving construction around $T_c$) is found to be approximately 1.51. This value exceeds the weak-coupling Bardeen-Cooper-Schrieffer (BCS) prediction of 1.43, suggesting that FeSe falls within an intermediate coupling strength rather than the weak-coupling limit. For single crystals with $x$ = 0.006, 0.014, 0.017, and 0.023, the electronic specific heat coefficient in the normal state, the fitted phonon parameters, and the corresponding Debye temperatures are summarized in Table II. In addition, the relatively low $\Theta_D$ extracted from the fits likely arise from the limited fitting range above $T_c$, where higher-order phonon contributions and additional low-energy excitations in FeSe deviate from the simple Debye $T^3$ approximation, leading to an overestimation of the $\beta$ coefficient.

**Table II**. Normal-state electronic specific-heat coefficient $\gamma_n$ (mJ·mol$^{-1}$·K$^{-2}$), fitted phonon expansion coefficients $\beta_1$ (mJ·mol$^{-1}$·K$^{-4}$), $\beta_3$ (mJ·mol$^{-1}$·K$^{-6}$), and $\beta_5$ (mJ·mol$^{-1}$·K$^{-8}$), $\Delta C/\gamma_n T_c$, and extracted Debye temperature $\Theta_D$ (K) for Fe$_{1-x}$Zn$_x$Se single crystals with $x$ = 0-0.023.

| Fe$_{1-x}$Zn$_x$Se | $\gamma_n$ | $\beta_1$ | $\beta_3$ | $\beta_5$ | $\Theta_D$ | $\Delta C/\gamma_n T_c$ |
|---|---|---|---|---|---|---|
| $x$ = 0 | 5.71 | 0.92 | -3.81×10$^{-3}$ | 9.07×10$^{-6}$ | 163 | 1.51 |
| $x$ = 0.006 | 5.80 | 1.01 | -4.23×10$^{-3}$ | 1.30×10$^{-5}$ | 158 | 1.41 |
| $x$ = 0.014 | 5.96 | 2.34 | -1.45×10$^{-2}$ | 4.47×10$^{-5}$ | 120 | 1.67 |
| $x$ = 0.017 | 5.08 | 0.80 | -2.92×10$^{-3}$ | 8.39×10$^{-6}$ | 171 | 1.76 |
| $x$ = 0.023 | 5.50 | 1.14 | -4.91×10$^{-3}$ | 1.33×10$^{-5}$ | 151 | 1.30 |

The electronic specific heat $C_e = C - C_n$ in the superconducting state was analyzed within the single-band isotropic $s$-wave $\alpha$-model, based on the BCS theory [44]. Within this framework, the normalized electronic specific heat is given by

$$\frac{C_e}{\gamma_n T} = \frac{6}{\pi^2} \int_0^\infty \left(\frac{E'}{k_B T}\right)^2 f(E')[1 - f(E')]dE,$$

where $f(E') = [exp(E'/k_B T) + 1]^{-1}$ is the Fermi-Dirac distribution function, and $E' = \sqrt{E^2 + \Delta^2(T)}$ is the quasiparticle excitation energy. The temperature dependence of the superconducting gap is approximated by

$$\Delta(T) = \Delta_0 tanh\left[1.74\sqrt{T_c/T - 1}\right],$$

which provides an accurate interpolation to the BCS gap function over most of the temperature range [45]. $\Delta_0$ is the superconducting energy gap for isotropic $s$-wave. To incorporate gap anisotropy, the superconducting order parameter can be generalized as $\Delta(\theta, T) = \Delta(T)g(\theta)$, and the specific heat is evaluated by averaging over the Fermi surface. For a $d$-wave gap symmetry, the angular dependence is typically expressed as $\Delta(\theta, T) = \Delta(T)cos(2\theta)$, which leads to line nodes on the Fermi surface and a characteristic power-law behavior of the low-temperature specific heat [46]. For an anisotropic $s$-wave gap, a commonly used form is $\Delta(\theta, T) = \Delta(T)(1 + a\cos(4\theta))$, which is often referred to as an extended $s$-wave gap, since it preserves the overall $s$-wave symmetry while allowing angular modulation of the gap amplitude [47-49]. Here, $a$ quantifies the degree of anisotropy and the gap remains nodeless when $a < 1$. To describe multiband superconductivity, the total electronic specific heat can be written as a weighted sum of contributions from different bands, $C_e = \omega C_1 + (1 - \omega)C_2$, where $\omega$ represents the relative weight of each band. Within the two-gap $\alpha$-model, this becomes $C_e = \omega C(\Delta_1) + (1 - \omega)C(\Delta_2)$, where $\Delta_1$ and $\Delta_2$ correspond to superconducting gaps on different Fermi surface sheets [50,51]. Overall, these extensions of the $\alpha$-model provide a flexible framework for distinguishing between isotropic, anisotropic (extended) $s$-wave, and multigap superconducting states based on thermodynamic measurements.

We first analyzed the normalized electronic specific heat $C_e/\gamma_n T_c$ of FeSe using a single-band isotropic $s$-wave model. However, the resulting fit shows clear deviations from the experimental data, as shown in Fig. S7. We then attempted alternative gap symmetries, including $d$-wave, two-gap $s$-wave, and extended $s$-wave models. Despite systematic variation of the fitting parameters, none of these scenarios provides a satisfactory description of the data (Fig. S7). Given the multiband nature of FeSe, we further employed a two-gap model combining an extended $s$-wave gap with an isotropic $s$-wave component. This approach yields a significantly improved agreement with the experimental results. The corresponding fitting parameters are $2\Delta_s/k_B T_c = 2.0$ and $2\Delta_{es}/k_B T_c = 4.9$ ($a = 0.43$) with the relative weight of 45% and 55% (see Table III). For $Fe_{1-x}Zn_xSe$ single crystals with different Zn concentrations, the normalized electronic specific heat exhibits behavior similar to that of the parent compound FeSe. As shown in inset of Fig. 6(c-f), Fig. S7 and Table III, the experimental data show that, apart from the two-gap model combining an extended $s$-wave and an isotropic $s$-wave component, other pairing symmetries fail to provide a satisfactory description of the data.

**Table III**. Fitting parameters for the normalized electronic specific heat of $Fe_{1-x}Zn_xSe$ single crystals, obtained using a two-gap model combining isotropic $s$-wave ($\Delta_s$) and extended $s$-wave components ($\Delta_{es}$).

| $Fe_{1-x}Zn_xSe$ | $2\Delta_s/k_B T_c$ | $2\Delta_{es}/k_B T_c$ | $a$ | $s$-wave (weight) | E. $s$-wave (weight) |
|---|---|---|---|---|---|
| $x = 0$ | 2.0 | 4.9 | 0.43 | 45% | 55% |
| $x = 0.006$ | 1.8 | 5.0 | 0.41 | 45% | 55% |
| $x = 0.014$ | 1.6 | 6.3 | 0.40 | 45% | 55% |
| $x = 0.017$ | 1.8 | 4.8 | 0.50 | 40% | 60% |
| $x = 0.023$ | 1.5 | 4.5 | 0.90 | 40% | 60% |

Furthermore, our results demonstrate that multigap superconductivity persists robust in Zn-substituted FeSe. The extracted gap magnitudes define two distinct energy scales, consistent with the multiband nature of superconductivity reported previously. Notably, the near-invariance of the relative weights associate with the two gap components upon Zn substitution suggest weak interband scattering remains weak, thereby preserving the multiband character of the superconducting state. Combined with the absence of rapid suppression of superconductivity by nonmagnetic impurities, this behavior disfavors a simple sign-changing pairing state in the strong interband-scattering limit and instead points to a predominantly sign-preserving pairing state with anisotropic gap structure.

The nonmonotonic evolution of the superconducting transition temperature $T_c$ with Zn substitution provides further insight into the underlying physical mechanism. At low Zn concentrations, enhanced impurity scattering suppresses superconductivity, while at intermediate doping levels, modifications to the relative contributions of different Fermi surface pockets may lead to a partial recovery of $T_c$. With further increase in Zn content, disorder and scattering effects dominate again, resulting in a renewed suppression of superconductivity. Such behavior deviates from the monotonic pair-breaking expected for sign-changing pairing states, such as *d*-wave or $s\pm$, under nonmagnetic impurity scattering, and instead reflects a competition between impurity scattering and multiband electronic effects.

Moreover, combining our experimental findings with theoretical considerations [52-54], we infer that FeSe likely lies in a crossover regime where $s\pm$ and $s++$ pairing tendencies compete. However, the ground state appears to be predominantly sign-preserving *s*-wave-like superconductivity with a substantial anisotropic component. In this scenario, Zn substitution primarily tunes scattering channels and the relative contributions of different Fermi-surface pockets, leading to the observed nonmonotonic $T_c$ evolution rather than simple pair-breaking behavior. These results place strong constraints on the pairing symmetry in FeSe and support a unified picture of multiband superconductivity characterized by competition between multiple pairing channels and weak interband scattering, emphasizing the material-dependent nature and many-body complexity of pairing mechanisms in iron-based superconductors [52-54].

**Conclusion**

In this work, we systematically investigated the superconductivity of $Fe_{1-x}Zn_xSe$ single crystals. Throughout the entire doping range, the samples maintain high crystalline quality and consistently exhibit robust bulk superconductivity, indicating that Zn substitution does not destroy the overall superconducting ground state. Notably, with increasing Zn content, $T_c$ exhibits a pronounced nonmonotonic behavior, indicating that its evolution cannot be explained by simple monotonic suppression due to impurity scattering, but instead arises from a competition between impurity scattering and multiband electronic effects. The low-temperature specific heat can be well described by a two-gap model consisting of an anisotropic extended *s*-wave gap and an isotropic *s*-wave gap, whereas single-gap and *d*-wave models fail to reproduce the experimental data. In addition, the nearly unchanged relative weights of the two gap components across different Zn concentrations indicate weak interband scattering induced by Zn substitution, thereby preserving the robustness of the multiband superconducting state, with possible implications for a sign-preserving pairing symmetry.

**Acknowledgement**

We sincerely appreciate Dr. Wenjie Li (Southeast University) for fruitful discussions, and we also thank the staff members of the Physical Property Measurement System (https://cstr.cn/31125.02.SHMFFPPMS) at the Steady High Magnetic Field Facility, CAS (https://cstr.cn/31125.02.SHMFF), for providing technical support and assistance in data collection and analysis. This work was supported by the National Natural Science Foundation of China (Grants

No. 12204452, 21671182), the National Key R&D Program of China (2021YFB4001401), the Grants for Scientific Research of BSKY from Anhui Medical University, and Guangzhou Higher Education Teaching Quality and Teaching Reform Project (2024YBJG024).

**Author declarations**

**Conflict of interest**

The authors declare no conflicts of interest.

**Data availability**

The data that support the findings of this study are available from the corresponding author upon reasonable request.

**Corresponding authors**

xhs@ustc.edu.cn
cohenkeyuan@hkust-gz.edu.cn
kbtang@ustc.edu.cn
yyhan@hmfl.ac.cn